\documentclass[12pt,reqno]{amsproc}
\usepackage{amsthm,amsfonts,latexsym,amssymb,
mathbbol,amsmath,
}
\usepackage[citation-order,short-journals]{amsrefs}
\usepackage[dvips]{graphics}
\newcommand{\be}{\begin{equation}}
\newcommand{\ee}{\end{equation}}
\newcommand{\bea}{\begin{eqnarray}}
\newcommand{\eea}{\end{eqnarray}}
\newcommand{\beax}{\begin{eqnarray*}}
\newcommand{\eeax}{\end{eqnarray*}}

\newcommand{\Ir}{\mathbb{Z}}
\newcommand{\Rl}{\mathbb{R}}

\numberwithin{equation}{section}
\newcounter{saveeqn}

\allowdisplaybreaks[3]
 
\newcommand{\N}{\mathbb{N}}
\newcommand{\Z}{\mathbb{Z}}
\newcommand{\R}{\mathbb{R}}
\newcommand{\C}{\mathbb{C}}
\newcommand{\I}{\Eins}
\newcommand{\J}{\mathcal{J}}
\newcommand{\A}{\ensuremath{\mathfrak{A}}}
\newcommand{\AL}{\ensuremath{\mathfrak{A}_{\text{loc}}}}
\newcommand{\Hi}[1][J]{{\mathcal{H}_{#1}}}

\newcommand{\xxz}{\textrm{XXZ} }
\newcommand{\ccr}{\textrm{CCR} }
\newcommand{\hlf}[1][1]{\scriptscriptstyle{#1/2}}
\newcommand{\thrd}[1][1]{\scriptscriptstyle{#1/3}}
\newcommand{\Stot}[1][,\Lambda]{\ensuremath{S^3_{\mathrm{tot}#1}}}
\newcommand{\Pszl}[1][z]{\ensuremath{\Psi_\Lambda^{(#1)}}}
\newcommand{\Fml}[1][M]{\ensuremath{\Phi_\Lambda^{(#1)}}}
\newcommand{\om}[1][M]{\ensuremath{\omega^{(#1)}}}

\newcommand{\thet}[1][r]{\ensuremath{\theta^{(#1)}}}
\newcommand{\hr}[1][r]{\ensuremath{\tilde h^{(#1)}}}
\newcommand{\Hr}[1][r]{\ensuremath{\tilde H^{(#1)}}}
\newcommand{\ux}[1][x]{\ensuremath{u^{(r)}_{#1}}}
\newcommand{\omrl}[1][r]{\ensuremath{\omega^{(#1)}_\Lambda}}
\newcommand{\omr}[1][r]{\ensuremath{\omega^{(#1)}}}

\newcommand{\Ntot}{\ensuremath{N_{\mathrm{tot}}}}
\newcommand{\Hkin}[1][J]{\ensuremath{H_{#1,\mathrm{kin}}}}
\newcommand{\Hdyn}{\ensuremath{H_{J,\mathrm{dyn}}}}
\newcommand{\Htran}{\ensuremath{H_{J,\mathrm{tran}}}}
\newcommand{\gapr}{\ensuremath{\tilde\gamma^{(r)}}}

\theoremstyle{plain}
\newtheorem{theorem}{Theorem}[section]

\newtheorem{proposition}[theorem]{Proposition}
\newtheorem{lemma}[theorem]{Lemma}
\newtheorem{corollary}[theorem]{Corollary}

\newtheorem{conjecture}[theorem]{Conjecture}
\newtheorem*{theorem*}{Theorem}
\theoremstyle{definition}
\newtheorem{remark}[theorem]{Remark}

\DeclareMathOperator*{\slim}{s-lim}
\DeclareMathOperator*{\otim}{\otimes}
\DeclareMathOperator*{\opl}{\oplus}

\DeclareMathOperator{\sgn}{sgn}

\DeclareMathSymbol{\defeq}{\mathrel}{symbols}{205}

\renewcommand{\phi}{\varphi}
\renewcommand{\epsilon}{\varepsilon}

\begin{document}

\title[Large-spin ferromagnetic XXZ chain]{The large-spin asymptotics 
of the ferromagnetic XXZ chain}
\author{Tom Michoel}
\address{Instituut voor Theoretische Fysica,
Katholieke Universiteit Leuven,
Celestijnenlaan 200 D,
B--3001 Leuven,  Belgium}
\email{tomm@itf.fys.kuleuven.ac.be}
\author{Bruno Nachtergaele}
\address{Department of Mathematics, University of California, Davis, 
One Shields Avenue, Davis 95616-8366, USA}
\email{bxn@math.ucdavis.edu}
\thanks{This material is based on work supported by the National Science
Foundation under Grant No.~DMS0303316. T. Michoel
is a Postdoctoral Fellow of the Fund for
Scientific Research -- Flanders (Belgium) (F.W.O.--Vlaanderen)}
\thanks{\copyright\ 2003 by the authors. This article may be reproduced in its
entirety for non-commercial purposes.}

\date{April 15, 2003}


\begin{abstract} 
We present new results and give a concise review of recent previous results
on the asymptotics for large spin of the low-lying spectrum of the
ferromagnetic XXZ Heisenberg chain with kink boundary conditions. Our main
interest is to gain detailed information on the interface ground states of
this model and the low-lying excitations above them. The new and most
detailed results are obtained using a rigorous version of bosonization,
which can be interpreted as a quantum central limit theorem.
\end{abstract}

\maketitle

\section{Introduction}

In recent years the XXZ model has become a popular model to study properties of
interface states in quantum lattice models. As an interpolation between the
Ising model and the isotropic (XXX) Heisenberg ferromagnet, the ferromagnetic
XXZ model has the interesting features of both. By considering the Ising model
and the XXX model as limiting cases of the XXZ model, intuition about these two
limits can be used to better understand the XXZ model. In this paper we are
interested in the large-spin asymptotics of the low-lying excitation spectrum
of the XXZ chain, in particular the excitations above the kink (or interface)
ground states of the model. In a nutshell, our main result is that the
spin-wave approximation, in the sense of Dyson \cite{dyson:1956}, becomes exact
in the limit of infinitely large spin. The technical statements are given in
Section \ref{sec:main_results}. First, we introduce the model and the main
notations and give a quick summary of the relevant previous results.

For $J=1/2,1,3/2,\ldots$, the spin-$J$ XXZ Hamiltonian on an interval 
$\Lambda=[a,b]\subset \Ir$, with kink boundary conditions, is given by
\begin{align}
  H_{J,\Lambda}&=\sum_{x=a}^{b-1} H^J_{x,x+1}\label{eq:1}\\
  H^J_{x,x+1}&=J^2-\frac 1\Delta(S^1_xS^1_{x+1}+ S^2_xS^2_{x+1}) 
  - S^3_xS^3_{x+1} + J\sqrt{1-\Delta^{-2}}(S^3_x-S^3_{x+1}) \nonumber
\end{align}
with $S^i_x$ the spin-$J$ matrices acting on site $x$:
\begin{align*}
  [S^i_x,S^j_y]&=i\delta_{x,y}\epsilon_{ijk} S^k_x\\
  S_x\cdot S_x&= (S_x^1)^2 + (S_x^2)^2 + (S_x^3)^2 = J(J+1)
\end{align*}
We will also use the spin raising and lowering operators:
$S^+_x$ and $S^-_x$, by $S^\pm_x=S^1_x\pm i S^2_x$.

We begin with a brief overview of the main results obtained for the 
Hamiltonians $H_{J,\Lambda}$. The spin 1/2 model, $J=1/2$, is Bethe  Ansatz
solvable and posseses a quantum group symmetry \cite{pasquier:1990}. 
Consequently, there are a number of results specific to the spin 1/2 case
(E.g., see \cite{koma:1997,bolina:2000,nachtergaele:2003}). Since the main
focus in this paper is on  large-$J$ behavior, we will not discuss these
specific results here.

The Hamiltonian \eqref{eq:1} is symmetric under global rotations around the
$3$-axis generated by $\Stot=\sum_{x\in\Lambda} S^3_x$, which represents
the third component of the total magnetization. Hence,
$H_{J,\Lambda}$ is block diagonal, and it is known that in each 
sector corresponding to a given eigenvalue of $\Stot$
there is exactly one ground state. i.e., in each sector $0$ is a simple
eigenvalue \cite{koma:1998}:
\begin{align*}
  H_{J,\Lambda}\Fml=0 && \Stot\Fml=M\Fml
\end{align*}
where $M=-|\Lambda|J,-|\Lambda|J+1,\ldots,|\Lambda|J$.
The (unnormalized) eigenvector $\Fml$ is given by
\begin{equation*}
  \Fml = \sum_{\substack{\{m_x\}\\\sum_x m_x=M}}
  \prod_{x\in\Lambda} \binom{2J}{J-m_x}^{1/2} q^{x(J-m_x)}
  |\{m_x\}\rangle
\end{equation*}
where we introduced the parameter $q$, $0<q<1$, by the equation
$2\Delta=q+q^{-1}$.
These ground states have a magnetization profile that shows an interface, 
or {\em kink}, with a location depending on the value of $M$. For a 
short review on the properties of these ground states see
\cite{nachtergaele:2001}.

In many instances it is important to consider the thermodynamic limit, i.e, 
the limit of infinitely long chains. To this end, we consider a strictly
increasing sequence of numbers $a_n\in\Z^+$ and $\lim_n a_n=\infty$, and a
sequence of volumes $\Lambda_n=[-a_n+1,a_n]$. The set of eigenvalues of
$S^3_{\text{tot},\Lambda_n}$ is then
\begin{equation*}
  \mathcal{M}_n=\{-2a_n J, -2a_nJ+1,\dots,2a_nJ\}
\end{equation*}
and since $2J$ is an integer, we have $\mathcal{M}_n\subset \mathcal{M}_m$ for
$n<m$. Hence we can fix $M\in\Z$, take $n_0$ large enough such that
$M\in\mathcal{M}_n$ for all $n\geq n_0$, and consider a sequence of states
\begin{equation*}
   \omega_{\Lambda_n}^{(M)}= \frac{\bigl\langle
     \Phi_{\Lambda_n}^{(M)}, \;\cdot\;
    \Phi_{\Lambda_n}^{(M)} \bigr\rangle}{\bigl\langle
    \Phi_{\Lambda_n}^{(M)},\Phi_{\Lambda_n}^{(M)} \bigr\rangle}
\end{equation*}

It is shown in \cite{gottstein:1995,koma:2001} that in the limit $n\to\infty$, any such
sequence converges in norm to a unique state $\om$ on the quasi-local algebra 
of observables $\A$ which is the norm completion of the algebra of local 
observables given by
\begin{equation*}
  \AL = \bigcup_{\Lambda\subset\Z}\otimes_{x\in\Lambda}\mathbb{M}_{2J+1}(\C)
\end{equation*}
and each $\om$ is a ground state for the derivation $\delta_J$
defined by
$$
  \delta_J(X)=\lim_{\Lambda\nearrow\Z}i[H_\Lambda^J,X] 
$$
i.e.,
$$ 
\om\bigl( X^*\delta(X)\bigr) \geq 0 \qquad \mbox{ for } 
  X\in\AL
$$

All these infinite volume kink states have the same GNS Hilbert space, 
namely the incomplete tensor product Hilbert space
\begin{equation*}
  \Hi=\overline{\bigcup_{\Lambda\subset \Z}\Bigl(
    \bigl[\otim_{x\in\Lambda} \C^{2J+1}\bigr] \otimes \bigl[
    \otim_{y\in\Z\setminus\Lambda}\Omega_y\bigr]\Bigr)}
\end{equation*}
where
\begin{equation*}
  \Omega_y=
  \begin{cases}
    |-J\rangle & \text{if $y\leq 0$}\\
    |J\rangle & \text{if $y> 0$}\\
  \end{cases}
\end{equation*}
Also denote $\Omega=\otim_{y\in\Z}\Omega_y \in\Hi$, and the GNS Hamiltonian on
$\Hi$ will be denoted $H_J$.

It was proved in \cite{starr:2001,koma:2001} that, for all
$J=1/2,1,3/2,\ldots$, these Hamiltonians have a gap above the ground state
eigenvalue, which is $0$. Let us denote the gap by $\gamma_{J,M}$. In the case
of $J=1/2$, the exact value of the gap  was previously known to be
$1-\Delta^{-1}$, for all $\Delta\geq 1$. In \cite{koma:2001} a very explicit
conjecture is made about the value of the gap in the limit $J\to \infty$.
For all finite $J$, it is a periodic function of $M$, with period $2J$. The
conjecture in \cite{koma:2001} is as follows: 

\begin{conjecture}
For all $\mu\in\Rl$, the limit
$$
\lim_{J\to\infty}\frac{1}{J}\gamma_{J,\mu J}=\tilde\gamma(r)
$$
exists and given by the smallest positive eigenvalue of the Jacobi operator
$\tilde{h}^{(r)}$, defined below in \eqref{jacobi}, where are is the
solution of the equation
$$
\mu=\sum_{x=-\infty}^{+\infty}\tanh(\eta(x-r))
$$
with $\eta$ determined by $\Delta=\cosh \eta$.
\end{conjecture}

A partial result towards this conjecture was proved in \cite{caputo:2003}.
Namely, there it is shown that there are constants $c_1>0$, and $c_2>0$,
independent of $M$ and $J$, such that 
$$
c_1 J\leq \gamma_{J,M}\leq c_2 J\ .
$$

In this paper we prove that the value of the gap claimed in the conjecture is
asymptotically correct. The results in this paper by themselves, however, do
not amount to a proof of the conjecture as stated. Roughly speaking, we obtain
the result in the ``grand-canonical ensemble'', and with the aid of a ground
state selection mechanism that localizes the kink. The conjecture is stated in
the ``canonical ensemble'', i.e., for fixed  $\mu=M/J$, in which the kink is
automatically localized at a fixed location. As is often the case the
distinction between canonical and grand-canonical results seems merely
technical, but proving mathematical equivalence of both formulations is often
highly non-trivial.  In fact, equivalence of ensembles in the usual sense does
not hold in the present situation. To prove the conjecture as stated above,
some additional work has to be done. We will report on this further work in a
future publication \cite{michoel:2003}. 

The conjecture of \cite{koma:2001} was based on results form perturbation
theory and numerical calculations on small systems presented in 
\cite{starr:2001}, as well as on a heuristic calculation leading to 
a Boson model. 

The idea is to apply a rigorous version of Dyson's spin wave formalism
to the XXZ chain. Mathematically speaking, the task is to control
the quadratic approximation, described by a quasi-free system of Bosons,
and show that this approximation becomes exact in the limit $J\to\infty$.

Several authors have attempted to do this for the XXX model, with interesting
results \cite{conlon:1990,hemmen:1984}. In these works, the authors considered
the XXX model in an external magnetic field, and it was necessary to let the
strength of the field diverge as $J\to\infty$. Such a field selects a
particular ground state (out of the infinite number of them), and creates a gap
in the spectrum. This allows one to proceed, but it limits the mathematical
applicability of the  spin-wave formalism. In the case of the XXZ-model, the
situation is somewhat better. First, the XXZ chain by itself (i.e., without
external magnetic field) already has a non-vanishing spectral gap. Second,
although the infinite XXZ chain also has an infinite number of ground
states---with the degeneracy now corresponding to the arbitrary position of the
kink---, any field at just one site with a non-vanishing component in the
XY-plane, a so-called pinning field, will select a unique ground state, for
finite  $J$ \cite{contucci:2002}. Moreover, the magnitude of this field, as we
will show, can be taken of smaller order in $J$. These features of the XXZ
model will allow us to prove asymptotic properties of the model itself.

\section{Main results}\label{sec:main_results}

\subsection{The limiting Boson model}\label{sec:limiting-boson}

Our main result will be that the spectrum of the XXZ chain, in the limit
of infinite spin, can be understood as the spectrum of a non-interacting
system of Bosons on the chain, with one-particle Hamiltonian,
$\tilde{h}^{(r)}$, defined on $\ell^2(\Ir)$ by
\begin{equation}
(\tilde{h}^{(r)}v)_x = \epsilon_x v_x -\frac1\Delta (v_{x-1}+ v_{x+1})
\label{jacobi}\end{equation}
where
\begin{equation*}
  \epsilon_x=
  \frac{2\cosh(\eta(x-r))^2}{\cosh(\eta(x-1-r)) \cosh(\eta(x+1-r))} 
\end{equation*}
with $\eta=-\ln q$ or, equivalently, $\Delta=\cosh \eta$, and $r\in\Rl$ is 
the position of the kink in the reference ground state.
$\tilde{h}^{(r)}$ has the form of the discrete Laplacian (kinetic energy) 
plus a diagonal term given by $\epsilon_x$, which is an exponentially 
localized potential well centered around the interface.

We list some properties of \hr, some of which are easily proved, while other
more detailed properties about its spectrum have at this point only been 
verified numerically. We will discuss these in more detail elsewhere.
\begin{enumerate}
\item \hr is a positive operator and hence the Fock state is a ground state for
  $\tilde\delta^{(r)}$; 
\item \hr has an eigenvalue $0$ with eigenvector $v_0\in l^2(\Z)$, up to
  normalization defined by
  \begin{equation*} 
    v_{0,x} = \sin\thet_x=\frac1{\cosh(\eta(x-r))}
  \end{equation*}
\item the bottom of the continuous spectrum of $\hr$ is given by $2(1-\Delta^{-1})$;
\item the first excited state of \hr corresponds to an isolated eigenvalue $\gapr$ below
  the continuum;
\item the $(\Delta^{-1},r)$-plane is divided in a region where \gapr is the only
  eigenvalue below the continuum, and a region where there is another isolated eigenvalue
  between \gapr and the bottom of the continuum.
\end{enumerate}

Let $\hr(x,y)$ be the bi-infinite Jacobi matrix expressing \eqref{jacobi}
in the standard Kronecker-delta basis of $\ell^2(\Ir)$, i.e.,
$$
\hr(x,y)=\epsilon_x\delta_{x,y}-\frac1\Delta (\delta_{x-1,y}+ \delta_{x+1,y})\ .
$$
The Boson Hamiltonian is then given by second quantization of $\hr$:
$$
\tilde{H}^{(r)}=\sum_{x,y\in \Ir}\hr(x,y) a^+_x a_y
$$
where $a^+_x$ and $a_y$ are the creation and annihilation operators
for a boson at site $x$ and $y$, respectively. They act on the 
bosonic Fock space with one-particle space $\ell^2(\Ir)$, $\mathcal{F}$, 
and satisfy the canonical commutation relations (CCR) 
$$
a_y a^+_x - a^+_x a_y =\delta_{x,y}\I, \quad
a_y a_x - a_x a_y = a^+_y a^+_x - a^+_x a^+_y = 0, \quad x,y\in\Ir
$$
Let $\tilde\Omega\in\mathcal{F}$ denote the vacuum vector which, 
up to a scalar factor is uniquely characterized by the property 
$a_x\tilde\Omega=0$, for all $x\in \Ir$.

We will often use the following standard orthonormal basis in
$\mathcal{F}$. Introduce 
\beax
\vec n &= \{ n_x\in\N\}_{x\in\Z}\nonumber\\
\mathcal{N} &= \Bigl\{ \vec n\mid \sum_x n_x <\infty \Bigr\}
\eeax
Then, the set $\{\phi_{\vec n}\mid \vec n \in \mathcal{N}\}$, 
where
\begin{equation}
  \phi_{\vec n} 
  = \prod_{x\in\Z} \frac1{(n_x!)^{\hlf}}  (a^*_x)^{n_x} \tilde\Omega
\end{equation}
is an orthonormal basis of $\mathcal{F}$.

The GNS Hilbert spaces of the spin chains, $\Hi$, $J\geq 1/2, 1,3/2,\ldots$,
can be identified with a nested sequence of subspaces of $\mathcal{F}$, defined
for each $J$, as the linear span of all vectors $\phi_{\vec n}$, with $n_x\leq
2J+1$. We will use this identification throughout the paper, and we will use
the same symbol $\phi_{\vec n}$ to denote a vector in the spin Hilbert space
$\Hi$ and the boson space $\mathcal{F}$. We will also use the
projections $P_{n,x}$ on $\mathcal{F}$ which projects onto the first $n$ 
boson states at site $x$, i.e., on the states $\phi_{\vec n}$ \eqref{eq:23} 
with $0\leq n_x\leq n$, and denote $P_n = \prod_x P_{n,x}$, i.e.,
$P_J$ projects onto the states $\phi_{\vec n}$ with $\vec n\in\mathcal{N}_J$, 
and hence
\begin{equation*}
  \Hi = P_J \mathcal{F}
\end{equation*}

More details on the Boson model are given in Section \ref{sec:notation-boson}.

\subsection{Ground state selection by the external field}

To prove full convergence of the low-energy spectrum, we need to add an external
field. A physical external field would be of the form
\be
\sum_x \vec h_{J,x}\cdot \vec S_x 
\label{eq:9}\ee
For our purposes, however, the field is a perturbation and our results will
generally be more interesting if we can proof them with smaller perturbations.
Ideally, a vanishingly small fied localized at one site shoudl suffice to
select a reference ground state. It turns out that we cannot quite do this
in the present setup. We shall use a perturbation of the form
\begin{equation}\label{eq:17}
  \sup_x\bigl(  \vec h_{J,x}\cdot \vec S_x \bigr)
\end{equation}
which is still significantly smaller than a uniform field. The meaning of this
operator is clear if we express states in a basis which is diagonal for each
$\vec h_{J,x}\cdot \vec S_x $. It is important to remark that we can take
\begin{equation*}
|\vec h_{J,x}| \equiv h_J = h (J\ln J)^{\thrd[2]}
\end{equation*}
such that after scaling with $J^{-1}$ the external field vanishes, 
in contrast with the fields employed in previous works on the XXX model
\cite{conlon:1990,hemmen:1984}. In fact some results can be obtained 
with $h_J\equiv0$, or $h_J= h\ln J$.

Mathematically, the field \eqref{eq:17} does slightly more than a field
localized at one site can achieve. It not only pins the interface, but also
puts some control on the local fluctuations around the selected ground state,
which cannot be controlled otherwise. The fact that we can let the field vanish
as $J$ increases, and that we do not need a global field like \eqref{eq:9},
are signs that these fluctuations are smaller in the XXZ model than in the
isotropic model.

There is another mechanism of selecting a ground state, namely by restricting
the full Hilbert space to a subspace of states with fixed total magnetization
in the $3$-direction, since it is known that in each such sector there is
exactly one ground state \cite{koma:1998}. The ground states that are pinned by
an external field are like grand-canonical averages of the canonical ground
states with fixed magnetization. In the limit $J\to\infty$ the canonical
description can be obtained from the grand canonical one by a result analogous
to the result of \cite[Section 5.11, 5.12]{starr:2001} about equivalence of
ensembles in the $2$-dimensional, spin-$\frac12$ XXZ model. Note however that
there is \emph{no} equivalence of ensembles in our situation, in the sense that
correlation functions are typically different. But the difference between a
canonical state and a grand-canonical state with the right average
magnetization can be expressed completely in terms of the fluctuations of the
total $3$-magnetization in the grand canonical state which are non-zero even in
the limit $J\to\infty$, while they are identically zero in the canonical state.
The results about the canonical description require significant additional work
and will be discussed elsewhere \cite{michoel:2003}.

%

Finally we mention that because of the pinning field \eqref{eq:17}, our results
give only a partial proof of \cite[Conjecture 2.5]{koma:2001}. A full proof
requires in addition that the lowest excited state can be obtained from the
ground state by acting on it with $1$ spin wave operator, or more generally by
a finite number of operators independent of $J$. This is a problem that
should be handled at the level of the spin system, rather than in the
spin wave formalism. Some result in that direction have recently been
obtained in \cite{nachtergaele:2003}.

An advantage of the grand-canonical description is that it clearly exhibits how the boson
limit arises as the first quantum correction to the classical limit.

\subsection{Statement of the main results}\label{sec:statements}

For $A$ a self-adjoint operator and $a,b\in\R$, denote by $P_{(a,b)}(A)$ the spectral
projection of $A$ onto $(a,b)$. For $A$ acting on Fock space, denote
\begin{itemize}
\item $\sigma(A)$ the spectrum of $A$ in $\mathcal{F}$;
\item $\sigma_J(A)$ the spectrum of $P_JAP_J$ in $\Hi$;
\item $\sigma_{n_J}(A)$ the spectrum of $P_{n_J}AP_{n_J}$ in $\Hi[n_J]$,
where $\Hi[n_J]$ is shorthand for $P_{n_J}\mathcal{F}$.
\end{itemize}
Also denote $\slim$ the strong, or strong resolvent, operator limit for bounded,
resp. unbounded operators acting on $\mathcal{F}$.

In the GNS space $\Hi$ it is convenient to define $\Stot[]$ in the renormalized sense: 
\begin{equation*}
  \Stot[] = \sum_{x\in\Z}\bigl[ S^3_x - \sgn(x-\tfrac12)\bigr]
\end{equation*}
Also denote
\begin{equation*}
  \mu=\sum_{x\in\Z} \bigl[\cos\thet_x - \sgn(x-\tfrac12)\bigr]
\end{equation*}
i.e., $\mu$ is the $3$-magnetization of a classical ground state
$\{(\thet_x,\phi)\}_{x\in\Z}$ (see section \ref{sec:xxz-class-lim}).

\begin{theorem}
  We have
  \begin{align*}
    \slim_{J\to\infty} \tfrac1J H_J &= \Hr\\
    \slim_{J\to\infty} \tfrac1J \Stot[] &= \mu \I
  \end{align*}
\end{theorem}

The proof of this result is given in Section \ref{sec:strong-conv}.

\begin{corollary}
  \begin{enumerate}
  \item If $\lambda\in\sigma(\Hr )$, there exists $\lambda_J\in\sigma_J(\tfrac1JH_J)$ such
    that
    \begin{equation*}
      \lim_{J\to\infty} \lambda_J = \lambda
    \end{equation*}
  \item If $a,b\in\R$, and $a,b\notin\sigma_{pp}(\Hr )$, then 
    \begin{equation*}
      \slim_{J\to\infty}  P_{(a,b)}(\tfrac1J H_J) =  P_{(a,b)}(\Hr)
    \end{equation*}
  \item If $a,b\in\R$, and $\mu\in (a,b)$, then
    \begin{equation*}
      \slim_{J\to\infty} P_{(a,b)}(\tfrac1J\Stot[]) = \I
    \end{equation*}
  \end{enumerate}
\end{corollary}

This corollary is a direct consequence of 
Proposition \ref{pro:strong-convergence} and 
\cite[Theorem VIII.24]{reed:1972}.

In addition we can obtain spectral concentration of $\tfrac1J H_J$ around
discrete eigenvalues of $\Hr $ (also proved in Section \ref{sec:strong-conv}).

\begin{theorem}
  For every isolated eigenvalue $E$ of $\Hr $, there exists an interval
  \begin{align*}
    I_J=(E-\epsilon_J, E+\epsilon_J)\quad\text{ with }\quad
    \lim_{J\to\infty} \epsilon_J\frac{J}{\ln J} = 0
  \end{align*}
  such that for any interval $I$ around $E$ s.t. $I\cap \sigma(\Hr) = \{E\}$:
  \begin{align*}
    \slim_{J\to\infty} P_{I\setminus I_J}\bigl(\tfrac 1J H_J\bigr) &= 0\\
    \slim_{J\to\infty} P_{I_J} \bigl(\tfrac 1J H_J\bigr)
    &= P_{\{E\}} \bigl(\Hr  \bigr)
  \end{align*}
\end{theorem}

Applying the same reasoning to $\Stot[]$, we find that the interval $(a,b)$ in item
  (iii) of Corollary \ref{cor:strong-convergence} can be chosen as
  \begin{equation*}
    (a,b) =  (\mu-\epsilon_J, \mu+\epsilon_J)
  \end{equation*}
  with again $\lim_J \epsilon_J J(\ln J)^{-1}=0$.

To prove full convergence of the spectrum, we have to add the external field
\eqref{eq:17} to $H_J$, or, to have a positive operator, add:
\begin{equation}
  F_J=\sup_x \bigl( |h_{J,x}| J - \vec h_{J,x}\cdot \vec S_x \bigr)
  = h_J \sup_x (J-\tilde S^3_x) = h_J \sup_x (N_x)
\end{equation}
with $h_J>0$, and $N_x=a^*_x a_x$. Let us
assume we add this field to $\tfrac 1J H^J$, so $h_J$ already contains the factor
$J^{-1}$.

Take $0<n_J<J$ as before, we get that on $\Hi[n_J]^\perp\cap\Hi$
\begin{equation*}
  \tfrac 1J H_J + h_J \sup_x N_x \geq h_J \sup_x N_x \geq h_J n_J \I
\end{equation*}
Clearly, by choosing $h_J$ such that
\begin{equation*}
  \lim_{J\to\infty} h_J n_J =\infty
\end{equation*}
statements about the spectrum on $\Hi$ reduce to statements about the spectrum on
$\Hi[n_J]$. Or, if one chooses to make statements about the spectrum below a certain value
$E$, it is sufficient to choose $h_J$ such that $\lim_J h_J n_J > E+\epsilon$.

\begin{theorem}
  Let $n_J = [ (J (\ln J)^{-1})^{\thrd}]$.  If $\lambda\notin \sigma(\Hr )$, then
  $\lambda\notin \sigma_{n_J}(\tfrac 1J H_J)$ for $J$ large enough.
\end{theorem}

This result is proved in Section \ref{sec:with-pinning}.

Hence we get convergence of the spectrum of $\tfrac 1J H_J + h_J \sup_x N_x$ if
\begin{align*}
  \lim_{J\to\infty} h_J =0 && \lim_{J\to\infty} h_J\Bigl(\frac{J}{\ln J}\Bigr)^{\thrd} =\infty
\end{align*}

\section{Derivation of the Boson limit}

\subsection{The classical limit}
\label{sec:xxz-class-lim}

The Boson limit can be considered as the first quantum correction the 
the classical limit. Therefore, we first discuss the classical limit.

It is well known that for any quantum spin system, after rescaling each spin
matrix by $J^{-1}$ and taking the large spin limit $J\to\infty$, one obtains
the corresponding classical spin system \cite{lieb:1973}. For the \xxz chain in
a finite volume this is defined by the Hamiltonian
\begin{equation}\label{eq:2}
  H_\Lambda^{\mathrm{cl}}\bigl(\{\sigma_x\}_{x\in\Lambda}\bigr) =
  \sum_{x=a}^{b-1}  1-\frac 1\Delta(\sigma^1_x\sigma^1_{x+1}+ 
  \sigma^2_x\sigma^2_{x+1}) - \sigma^3_x\sigma^3_{x+1} +
  \sqrt{1-\Delta^{-2}}(\sigma^3_x-\sigma^3_{x+1})
\end{equation}
where $\sigma_x$ is a unit vector in $\R^3$. Minimizing this function with respect to
$\{\sigma_x\}_x$ yields zero-energy configurations that are plane waves \cite{starr:2001},
i.e., in spherical coordinates we find configurations
$\sigma^{(r)}_x(\phi)=(\thet_x,\phi)$, $\phi\in[-\pi,\pi]$ (the same at all sites), and
\begin{equation*}
  \thet_x = 2\arctan(q^{x-r})
\end{equation*}
where $r\in\R$ determines the value of the total $3$-magnetization.

Defining $\eta=-\ln q$ or $\Delta=\cosh(\eta)$, we have
\begin{align}
  \cos\thet_x &= \frac{1-q^{2(x-r)}}{1+ q^{2(x-r)}} =
  \tanh\bigl(\eta(x-r)\bigr) \label{eq:21}\\
  \sin\thet_x &= \frac{2q^{x-r}}{1+q^{2(x-r)}}= \frac{1}{\cosh\bigl(\eta (x-r)\bigr)}\label{eq:22}
\end{align}
such that the zero-energy solutions clearly describe kinks centered around $r$.

For the classical model, to look at the low-energy behavior amounts to making a quadratic
Taylor approximation to \eqref{eq:2}. At each site, the angle coordinates are replaced by
new coordinates
\begin{align*}
  q_x = \theta_x - \thet_x && p_x = \sin\thet_x\;(\phi_x-\phi)
\end{align*}
and the resulting harmonic oscillator Hamiltonian is
\bea
  \tilde H_\Lambda^{\mathrm{cl}}&=&\frac12 \sum_{x=a}^{b-1}\Bigl( \epsilon_x^+ 
  (q_x^2 + p_x^2) - \frac1\Delta(q_xq_{x+1}+p_xp_{x+1}) \Bigr)\nonumber\\
  &&+\frac12 \sum_{x=a+1}^{b}\Bigl( \epsilon_x^- 
  (q_x^2 + p_x^2) - \frac1\Delta(q_xq_{x-1}+p_xp_{x-1}) \Bigr)\label{eq:3}
\eea
where $\epsilon_x^\pm$ are given by
\begin{equation*}
  \epsilon^\pm_x = \frac{\sin\thet_{x\pm 1}}{\Delta\sin\thet_x} =
  \frac{\cosh(\eta(x-r))}{\Delta\cosh(\eta(x\pm 1-r))} 
\end{equation*}
This can be derived using the identities of Lemma \ref{lem:angles} below.

At sites other than the boundary sites we have a single-site potential
\begin{equation*}
  \epsilon_x=\epsilon^+_x + \epsilon^-_x=
  \frac{2\cosh(\eta(x-r))^2}{\cosh(\eta(x-1-r)) \cosh(\eta(x+1-r))} 
\end{equation*}
which is an exponentially localized well centered around the interface.

\subsection{Grand canonical states}
\label{sec:gc-states}

For $(\theta_x,\phi_x)$ a general unit vector on the sphere at site $x$, we can define the
coherent spin state in $(\C^{2J+1})_x$ (see \cite{arecchi:1972,lieb:1973}):
\begin{align*}
  |(\theta_x,\phi_x)\rangle &= e^{\frac12\theta_x( S^-_x e^{i\phi_x} -
    S^+_x e^{i\phi_x})}|J\rangle\\
  &= \sum_{m_x=-J}^J \binom{2J}{J-m_x}^{1/2}(\cos\tfrac12\theta_x)^{J+m_x}
  (\sin\tfrac12\theta_x)^{J-m_x} e^{i(J-m_x)\phi_x}\;|m_x\rangle
\end{align*}

This is particularly interesting if we choose the unit vectors at each site to
be the classical zero-energy configurations.  In a finite volume $\Lambda$, it
is easy to see that
\begin{equation*}
  |\sigma_\Lambda^{(r)}(\phi)\rangle\equiv
  \otim_{x\in\Lambda}e^{-iJ\phi}|(\thet_x,\phi)\rangle = 
  \frac{1}{\|\Pszl\|}\Pszl 
\end{equation*}
where \Pszl is the generating vector for the ground state vectors \Fml, i.e., the grand
canonical ground state:
\begin{equation*}
  \Pszl = \sum_{M=-|\Lambda|J}^{|\Lambda|J} z^M \Fml
\end{equation*}
evaluated at $z=q^re^{-i\phi}=e^{-\eta r}e^{-i\phi}$.

The fact that a classical ground state yields an exact quantum ground state
through the coherent state representation, is because $H_{J,\Lambda}$ is a
normal Hamiltonian in the sense of \cite{lieb:1973}, and the classical and
quantum ground state energies are (exactly) related by the scaling factor $J^2$.

Since these states are product states, their thermodynamic limit is easily obtained. In
the GNS Hilbert space $\Hi$, define the embedding of $|\sigma_\Lambda^{(r)}(\phi)\rangle$
as
\begin{equation*}
  |\bar{\sigma}_\Lambda^{(r)}(\phi)\rangle \equiv
  |\sigma_\Lambda^{(r)}(\phi)\rangle \otimes \bigl[
  \otim_{x\in\Z\setminus   \Lambda} \Omega_x\bigr]
\end{equation*}

\begin{lemma}
For a sequence of intervals $\Lambda_n=[-a_n+1,a_n]$ 
tending to $\Z$, we have for $m>n$
  \begin{equation*}
    \bigl\| |\bar{\sigma}_{\Lambda_m}^{(r)}(\phi)\rangle -
    |\bar{\sigma}_{\Lambda_n}^{(r)}(\phi)\rangle \bigr\| \leq 
    2J  q^{2a_n}\frac{1-q^{2(a_m-a_n)}}{1-q^2} \bigl( q^{2r}+
    q^{2-2r} \bigr)
  \end{equation*}
\end{lemma}
\begin{proof}
  \begin{align*}
    &\bigl\| |\bar{\sigma}_{\Lambda_m}^{(r)}(\phi)\rangle -
    |\bar{\sigma}_{\Lambda_n}^{(r)}(\phi)\rangle \bigr\|^2 = \bigl\|
    \otim_{x\in\Lambda_m\setminus\Lambda_n} e^{-iJ\phi} |(\thet_x,\phi)\rangle -
    \otim_{x\in\Lambda_m\setminus\Lambda_n}
    \Omega_x \bigr\|^2\\
    &\quad= 2 - \prod_{x=-a_m+1}^{-a_n} e^{-iJ\phi} \langle -J |(\thet_x,\phi)\rangle
    \prod_{x=a_n+1}^{a_m} e^{-iJ\phi} \langle
    J|(\thet_x,\phi)\rangle\\
    &\qquad\qquad- \prod_{x=-a_m+1}^{-a_n}e^{iJ\phi} \langle (\thet_x,\phi)| -J\rangle
    \prod_{x=a_n+1}^{a_m} e^{iJ\phi}\langle
    (\thet_x,\phi)|J\rangle\\
    &\quad= 2-2\prod_{x=a_n}^{a_m-1}\frac1{(1+q^{2(x+r)})^J}
    \prod_{x=a_n+1}^{a_m} \frac1{(1+q^{2(x-r)})^J} \\
    &\quad\leq 2\Bigl( 1 - e^{-J\sum_{x=a_n}^{a_m-1}q^{2(x+r)}-J
      \sum_{x=a_n+1}^{a_m}q^{2(x-r)}}  \Bigr)\\
    &\quad\leq 2J\Bigl( \sum_{x=a_n}^{a_m-1}q^{2(x+r)} +
    \sum_{x=a_n+1}^{a_m}q^{2(x-r)}  \Bigr)\\
    &\quad= 2J q^{2a_n}\frac{1-q^{2(a_m-a_n)}}{1-q^2} \bigl( q^{2r}+ q^{2-2r} \bigr)
  \end{align*}
  where we used the inequalities (for $u\geq 0$) $1+u\leq e^u$ and $1-e^{-u}\leq u$.
\end{proof}
It follows that the sequence $|\bar{\sigma}_{\Lambda_n}^{(r)}(\phi)\rangle$ has a limit
$|\sigma^{(r)}(\phi)\rangle$ in $\Hi$ that we can formally write as
\begin{equation*}
  |\sigma^{(r)}(\phi)\rangle\equiv \bigl[ \otim_{x\leq 0} e^{-\frac12 (\pi
    -  \thet_x) ( S^-_x e^{i\phi} -  S^+_x e^{i\phi})}\Omega_x \bigr]
  \otimes \bigl[ \otim_{x>0} e^{\frac12\thet_x ( S^-_x e^{i\phi} -
    S^+_x e^{i\phi})}\Omega_x \bigr] 
\end{equation*}

In general, if we write $z=|z|e^{-i\phi}$, then
\begin{equation}\label{eq:18}
  \Pszl = e^{-i\phi\Stot}\Pszl[|z|]
\end{equation}
so it will be sufficient to restrict our detailed analysis to the 
grand canonical states $\Pszl[e^{-\eta r}]$. The expectation in this state 
will be denoted \omrl:
\begin{equation*}
  \omrl = \frac{\bigl\langle\Pszl[e^{-\eta r}], \;\cdot\;
    \Pszl[e^{-\eta r}]\bigr\rangle}{\bigl\|\Pszl[e^{-\eta r}]\bigr\|^2} 
\end{equation*}
and its thermodynamic limit $\omr$.

In this case, the coherent states are rotations of the `top' state $|J\rangle$ through an
angle $\thet_x$ around the $2$-axis,
\begin{equation*}
  |(\thet_x,0)\rangle = e^{-i\thet_xS^2_x}|J\rangle
\end{equation*}
Introduce the notation $\ux=(\thet_x,0)$, or in Cartesian coordinates
\begin{equation*}
  \ux=(\sin\thet_x,0,\cos\thet_x)
\end{equation*}

In the remainder, we will always keep $r$ fixed and do not make explicit the dependence on
$r$ of various quantities. Notice that by periodicity it is sufficient to take
$r\in[0,1)$.

Denote by $\{e^1_x,e^2_x,e^3_x\}$ the standard basis in $\R^3$ (the same at every
site), and
\begin{align*}
  f^1_x &=\cos\thet_x\;e^1_x -\sin\thet_x\;e^3_x\\
  f^2_x &= e^2_x\\
  f^3_x &= \ux = \sin\thet_x\;e^1_x+\cos\thet_x\; e^3_x\\
  \intertext{Conversely}
  e^1_x &= \cos\thet_x\;f^1_x +\sin\thet_x\;f^3_x\\
  e^2_x &= f^2_x\\
  e^3_x &= -\sin\thet_x\;f^1_x + \cos\thet_x\;f^3_x
\end{align*}
i.e., $\{f^1_x,f^2_x,f^3_x\}$ form an orthonormal frame for $\R^3$ and $\{f^1_x,f^2_x\}$
an orthonormal frame for the tangent plane $\R^2$ to the unit sphere at $\ux$.

For $v_x\in\R^3$, we denote by $\tilde v_x\in\R^2$ the projection of $v_x$ onto the
tangent plane at $\ux$ (shifted to the origin), i.e., $\tilde v_x=(\tilde v^1_x, \tilde
v^2_x)$ and
\begin{align*}
  \tilde v^1_x = v_x\cdot f^1_x = \cos\thet_x v^1_x - \sin\thet_x v^3_x && \tilde v^2_x
  = v_x\cdot e^2_x = v^2_x
\end{align*}
Conversely, if $\tilde v_x\in\R^2$ we associate to it a vector $v_x\in\R^3$ by putting the
component along the $\ux$-axis zero:
\begin{align*}
  v^1_x = \cos\thet_x \tilde v^1_x && v^2_x = \tilde v^2_x && v^3_x = -\sin\thet_x
  \tilde v^1_x
\end{align*}

Also denote
\begin{equation*}
  v_x\cdot S_x = v^1_x S^1_x + v^2_x S^2_x + v^3_x S^3_x
\end{equation*}
and define rotated spin operators (\cite[Eq. (3.9)]{arecchi:1972})
\begin{align}
  \tilde S^1_x &= e^{-i\thet_x S^2_x}S^1_x e^{i\thet_x S^2_x}= f^1_x\cdot S_x =
  \cos\thet_x S^1_x - \sin\thet_x
  S^3_x \label{eq:28}\\
  \tilde S^2_x &= e^{-i\thet_x S^2_x}S^2_xe^{i\thet_x S^2_x} = S^2_x
  \label{eq:29}\\
  \tilde S^3_x &= e^{-i\thet_x S^2_x}S^3_xe^{i\thet_x S^2_x} = f^3_x\cdot S_x =
  \sin\thet_x S^1_x + \cos\thet_x S^3_x \label{eq:30}
\end{align}
Hence we find that $|(\thet_x,0)\rangle$ is the `top' state for the rotated spin
operators:
\begin{equation}\label{eq:19}
  \tilde S^3_x|(\thet_x,0)\rangle = \tilde S^3_x e^{-i\thet_x S^2_x}
  |J\rangle = e^{-i\thet_x S^2_x} S^3_x |J\rangle = J
  |(\thet_x,0)\rangle 
\end{equation}
The rotated spin raising and lowering operators are
\begin{align*}
  \tilde S^\pm_x &= \tilde S^1_x\pm i\tilde S^2_x
  = -\sin\thet_x S^3_x + \cos\thet_x S^1_x \pm i S^2_x\\
  \intertext{or} \tilde S^+_x &= -\sin\thet_x S^3_x + \cos^2\frac{\thet_x}2\;
  S^+_x - \sin^2\frac{\thet_x}2\; S^-_x \\
  \tilde S^-_x &= -\sin\thet_x S^3_x - \sin^2\frac{\thet_x}2\; S^+_x +
  \cos^2\frac{\thet_x}2\; S^-_x
\end{align*}

One of the main observations is that because of \eqref{eq:19}, it is much more
convenient to introduce the spin wave formalism in the rotated spin basis than in the
original one. Following \cite{hemmen:1984}, introduce
\begin{align}
  \vec n &= \{ n_x\in\N\}_{x\in\Z}\nonumber\\
  \mathcal{N} &= \Bigl\{ \vec n\mid \sum_x n_x <\infty \Bigr\} \qquad
  \mathcal{N}_J= \Bigl\{ \vec n\mid \forall x\colon n_x\leq 2J ,
  \sum_x n_x <\infty \Bigr\}
  \label{eq:25}\\
  \phi_{\vec n} 
  &= \prod_{x\in\Z} \frac1{n_x!} \binom{2J}{n_x}^{-\frac12} \bigl(\tilde
  S^-_x\bigr)^{n_x} \Omega^{(r)}\label{eq:24}
\end{align}
The set $\{\phi_{\vec n}\mid \vec n\in\mathcal{N}_J\}$ 
is an orthonormal basis for $\Hi$.

We conclude with a little lemma that complements \eqref{eq:18}.
\begin{lemma}\label{lem:gc-state-rot}
  For $-\pi<\phi<\pi$,
  \begin{equation*}
    e^{i\phi\Stot} |\sigma_\Lambda^{(r)}\rangle =
    \Bigl(\cos(\tfrac12\phi) +i \cos\thet_x
    \sin(\tfrac12\phi)\Bigr)^{2J} e^{-i\sum_{x\in\Lambda}
      \alpha_x(\phi) \sin\thet_x \tilde S^-_x}
    |\sigma_\Lambda^{(r)}\rangle
  \end{equation*}
  where
  \begin{equation*}
    \alpha_x(\phi) = \frac{\sin(\tfrac12\phi)}{\cos(\tfrac12\phi) +
      i\cos\thet_x\sin(\tfrac12\phi)}  
  \end{equation*}
\end{lemma}
\begin{proof}
  We write the disentanglement relation \cite[Eq. (A4)]{arecchi:1972}
  \begin{equation*}
    e^{i\phi S^3_x} = e^{i\phi(-\frac12 \sin\thet_x (\tilde S^+_x
    +\tilde S^-_x) + \cos\thet_x \tilde S^3_x)} = e^{-iy_-\tilde
    S^-_x} e^{(\ln y_3) \tilde S^3_x} e^{iy_+ \tilde S^+_x}
  \end{equation*}
  where
  \begin{align*}
    y_3 &= \Bigl( \cos(\tfrac12\phi) + i\cos\thet_x\sin(\tfrac12\phi)
    \Bigr)^2 \\
    y_+ &= y_- = \frac{\sin\thet_x \sin(\tfrac12\phi)}{\cos(\tfrac12\phi) +
      i\cos\thet_x\sin(\tfrac12\phi)} = \alpha_x(\phi)\sin\thet_x
  \end{align*}
  and recall that $|\sigma_\Lambda^{(r)}\rangle$ is the product state of `top' states for
  the $\tilde S$-operators.
\end{proof}

It is now also clear how to choose the external field 
$\vec h_{J,x}$ in \eqref{eq:9} such that
\begin{equation*}
  \vec h_{J,x}\cdot\vec S_x = -h_J \tilde S^3_x\ ,
\end{equation*}namely
\begin{equation*}
  \vec h_{J,x} = -h_J u_x^{(r)}, \quad h_J>0\ .
\end{equation*}

\subsection{The boson chain}
\label{sec:notation-boson}

We consider immediately the infinite volume situation.  Consider the Hilbert
space of wave functions $l^2(\Z)$ which we alternatively consider as the usual
complex Hilbert space with inner product
\begin{equation*}
  \langle v,w\rangle = \sum_{x\in\Z}\overline{v}_xw_x
\end{equation*}
or as a real linear space with symplectic form $\sigma$ and complex structure $\J$, i.e.,
\begin{align*}
  v\in l^2(\Z)&=\bigl( (v^1_x,v^2_x)\in\R^2\bigr)_{x\in\Z}\\
  \sigma(v,w) &= \sum_{x\in\Z} v^1_x w^2_x - v^2_x w^1_x\\
  \J v &= \bigl( (-v^2_x,v^1_x)\in\R^2\bigr)_{x\in\Z}
\end{align*}
The \ccr-algebra $\ccr(l^2(\Z),\sigma)$ is generated by unitaries $\{W(v)\mid v\in
l^2(\Z)\}$ which satisfy the commutation relations
\begin{equation*}
  W(v) W(w)=e^{-\frac i2\sigma(v,w)} W(v+w)
\end{equation*}

The Fock state $\tilde\omega$ is the quasi-free state on $\ccr(l^2(\Z),\sigma)$ determined
by
\begin{equation*}
  \tilde\omega\bigl( W(v)\bigr) = e^{-\frac12 \langle v,v\rangle}
\end{equation*}
Its GNS representation is the usual Fock representation on a Fock space $\mathcal{F}$ with
a vacuum vector $\tilde\Omega = \otim_{x\in\Z}|0\rangle_x$ and creation and annihilation
operators $a^\sharp_x$ such that
\begin{equation*}
  W(v)= e^{i\sum_{x\in\Z}(v_xa^*_x + \overline{v}_x a_x)}
\end{equation*}
(We don't bother to distinguish between $W(v)$ and its representative in the Fock
representation).

In this Fock representation we can define a quasi-free Boson Hamiltonian by 
ca\-nonically quantizing the classical harmonic oscillator Hamiltonian 
\eqref{eq:3}. This means replacing the position and momentum variables
by canonical pairs $q_x$ , $p_x$, with commutation relations
$[q_x,p_y]=i\delta_{x,y}$ and
\begin{align*}
  a^*_x=\frac{q_x- ip_x}{\sqrt 2} && a_x=\frac{q_x+ ip_x}{\sqrt 2}
\end{align*}
The result is
\begin{align*}
  \tilde H_\Lambda^{(r)} &= \sum_{x=a}^{b-1}\Bigl( \epsilon_x^+ \bigl(a^*_x a_x +
  \frac12\bigr) - \frac 1\Delta a^*_x a_{x+1} \Bigr) + \sum_{x=a+1}^b \Bigl(
  \epsilon_x^- \bigl(a^*_x a_x+\frac12\bigr) - \frac 1\Delta a^*_x a_{x-1} \Bigr)
\end{align*}
The corresponding infinite volume derivation is denoted
$\tilde\delta^{(r)}(\cdot)=\lim_{\Lambda\nearrow \Z}i[\tilde H_\Lambda^{(r)},\cdot]$, and
the GNS Hamiltonian is denoted $\Hr $,
\begin{equation*}
  \Hr  = \sum_{x\in\Z} \epsilon_x a^*_x a_x - \Delta^{-1} a^*_x (a_{x-1}+ a_{x+1})
\end{equation*}

Denote $a^*(v)=\sum_x v_x a^*_x$. If $v$ is local, i.e., has only finitely many $v_x$
non-zero, then
\begin{equation*}
  \lim_{\Lambda\nearrow\Z}[\tilde H_\Lambda^{(r)}, a^*(v)] =
  a^*(\hr v)
\end{equation*}
where \hr is the bi-infinite Jacobi matrix defined on $l^2(\Z)$ by
\begin{equation}\label{eq:4}
  (\hr v)_x = \epsilon_x v_x -\frac1\Delta (v_{x-1}+ v_{x+1})
\end{equation}
For the finite system localized in $\Lambda=[a,b]$, we have $i[\tilde H_\Lambda^{(r)},
a^*(v)]=a^*(\hr_\Lambda v)$, where $(\hr_\Lambda v)_x=(\hr v)_x$ for $x$ in the bulk
$[a+1,b-1]$, and at the boundary sites we get
\begin{align*}
  \hr_\Lambda v_a &= \epsilon^+_a v_a - \frac1\Delta  v_{a+1}\\
  \hr_\Lambda v_b &= \epsilon^-_b v_b - \frac1\Delta v_{b-1}
\end{align*}

Some important properties of \hr were given in Section \ref{sec:limiting-boson}.
Recall the existence of a zero mode (property (ii)) given by:
\begin{equation*} 
    v_{0,x} = \sin\thet_x=\frac1{\cosh(\eta(x-r))}
\end{equation*}
That this is an eigenvector of \hr with eigenvalue zero, is easily verified:
\begin{align*}
    \epsilon_x^\pm v_{x,0} - \frac1\Delta v_{x\pm 1,0} =\frac{\sin\thet_{x\pm 1}}{\Delta
    \sin\thet_x} \sin\thet_x - \frac1\Delta \sin\thet_{x\pm1} =0
\end{align*}
The origin of the zero-mode $v_0$ is well understood. It arises from the rotation symmetry
of $H_{J,\Lambda}$. This can be seen from Lemma \ref{lem:gc-state-rot}, by taking
$\phi\propto J^{-\hlf}$ and formally identifying $J^{-\hlf} \tilde S^-_x$ with
$a^*_x$ (this identification will be made more precise below). For every $N\in\N$, there
is a zero-energy vector $\psi_{0,N}$ corresponding to an $N$-particle occupation of $v_0$:
\begin{equation*}
  \psi_{0,N} = \frac1{|v_0|_2^N\sqrt{N!}} a^*(v_0)^N \tilde\Omega
\end{equation*}
We define $P_0$ the projection on $v_0$ and $\tilde P_0$ the projection onto the
zero-energy vectors, i.e.,
\begin{align}
  P_0 &= \frac{|v_0\rangle\langle v_0|}{|v_0|_2}\label{eq:27}\\
  \tilde P_0 &= \opl_{N\in\N} |\psi_{0,N}\rangle \langle \psi_{0,N}|
\end{align}

For completeness we mention that we will always use the standard orthonormal
basis in $\mathcal{F}$, i.e., the set $\{\phi_{\vec n}\mid \vec n \in
\mathcal{N}\}$, where now
\begin{equation}\label{eq:23}
  \phi_{\vec n} = \prod_{x\in\Z} \frac1{(n_x!)^{\hlf}}  (a^*_x)^{n_x} \tilde\Omega
\end{equation}
We use the same symbol $\phi_{\vec n}$ to denote a vector in the spin Hilbert
space $\Hi$ and the boson space $\mathcal{F}$ since we will use the
identification of $\Hi$ with a subspace of $\mathcal{F}$ as discussed before.

\section{The large spin limit as a quantum central limit}
\label{sec:boson-lim-gc}

Since the grand canonical states can be written as the `all $+$' state for rotated spin
operators, we are in the usual situation of a fully ferromagnetic state in which we expect
a boson limit after rescaling with $J^{-\hlf}$, i.e., define
\begin{equation*}
  \frac1{\sqrt {2J}} \tilde S^-_x \to a^*_x
\end{equation*}
in some sense.

One way to make this precise is to define fluctuation operators: for
$v_x\in\R^3$,
\begin{equation*}
  F_J(v_x)=\frac1{\sqrt{J}} \bigl(
  v_x\cdot S_x - \omr(v_x\cdot S_x) \bigr)
\end{equation*}
i.e., $F_J(v_x)$ measures the deviation from the ground state expectation value of the
spin in the $v_x$ direction. Similar fluctuation operators are used to study fluctuations
of extensive observables, and their thermodynamic limit can be taken as a noncommutative
central limit \cites{goderis:1989b,goderis:1990}.

A connection between the spin limit $J\to\infty$ and these quantum central limits was made
in \cite{michoel:1999c}, with the caveat that each spin-$J$ had to be represented as a sum
of spin-$\frac12$'s, instead of working with an irreducible representation. This latter
restriction is however not necessary. We have, using results from
\citelist{\cite{arecchi:1972} \cite{lieb:1973}}:
\begin{align}
  \omr\bigl(e^{iv_x\cdot S_x}\bigr) 
  &= \Bigl\{ \cos(\frac12 |v_x|)+ i \frac{v_x\cdot \ux}{|v|}
    \sin (\frac12 |v_x|) \Bigr\}^{2J}
  \label{eq:6}\\
  \omr(v_x\cdot S_x) &= J (v_x\cdot \ux) \nonumber\\
  \omr\bigl( F_J(v_x)F_J(w_x) \bigr) &= \frac12\bigl( v_x\cdot w_x - (v_x\cdot
  \ux)(w_x\cdot \ux) + i (v_x\times w_x)\cdot \ux \bigr) \nonumber
\end{align}
The latter quantity defines a (degenerate) inner product on $\R^3$:
\begin{equation}\label{eq:5}
  \langle v_x,w_x\rangle_x = 2\omr\bigl( F_J(v_x)F_J(w_x) \bigr)
\end{equation}
It is not hard to use \eqref{eq:6} to show that
\begin{equation}\label{eq:20}
  \lim_{J\to\infty}\omr\bigl( e^{iF_J(v_x)} \bigr) = e^{-\frac12 \langle v_x,v_x\rangle}
\end{equation}

Clearly if either $v_x$ or $w_x$ is along the $\ux$-direction, then $\langle
v_x,w_x\rangle_x=0$. Hence \eqref{eq:5} defines an inner product in $\R^2=\C$, the tangent
plane to the unit sphere at $\ux$.  If for $\tilde v_x,\tilde w_x\in\R^2$, $v_x,w_x$ are
the corresponding vectors in $\R^3$ (see Section \ref{sec:gc-states}),
\begin{equation*}
  \langle \tilde v_x,\tilde w_x\rangle_x \equiv \langle
  v_x,w_x\rangle_x 
  =\overline{(\tilde v_x^1 + i\tilde v_x^2)}(\tilde w_x^1 + i\tilde
  w_x^2)
\end{equation*}
i.e., the standard inner product in $\C$.  We see that there are no fluctuations in the
direction perpendicular to the tangent plane at the classical zero-energy solution.  Two
vectors in $\R^3$ at the same site will be called equivalent if their projection onto the
tangent plane at $\ux$ is the same.

For $v=(v_x\in\R^3)_{x\in\Z}$, with only finitely many $v_x\not=0$, we simply extend this
by putting
\begin{align*}
  F_J(v) &= \sum_{x\in\Z} F_J(v_x)\\
  \langle v,w\rangle &= 2\omega\bigl( F_J(v)F_J(w)\bigr) =\sum_{x\in\Z} \langle
  v_x,w_x\rangle_x
\end{align*}
Using \eqref{eq:20}, and standard techniques, the quantum central limit theorem follows:
\begin{equation*}
  \lim_{J\to\infty}\omr\bigl( e^{iF_J(v_1)}\dots e^{iF_J(v_n)}\bigr) = \tilde\omega\bigl(
  W(v_1) \dots W(v_n)\bigr)
\end{equation*}
where $\tilde\omega$ is the Fock state on the \ccr-algebra $\ccr(l^2(\Z),\sigma)$
introduced in section \ref{sec:notation-boson}, and the vectors $v_1,\dots,v_n$ on the
r.h.s. mean their respective equivalence classes in $l^2(\Z)$.  The result can be
understood intuitively from
\begin{equation*}
  v_x\cdot S_x = \tilde v^1_x\tilde S^1_x + \tilde v^2_x\tilde S^2_x
  = (\tilde v^1_x-i\tilde v^2_x)\tilde S^+_x + (\tilde v^1_x+i\tilde
  v^2_x)\tilde S^-_x 
\end{equation*}
for $\tilde v_x\in\R^2$ and corresponding $v_x\in\R^3$.

When studying properties of the GNS Hamiltonian, it is actually easier to make a
correspondance between the GNS Hilbert spaces $\Hi$ of $\omr$ and $\mathcal{F}$ of
$\tilde\omega$. 

Following \cite{hemmen:1984}, introduce the projection $P_{J,x}$ on $\mathcal{F}$ which
projects onto the first $2J+1$ boson states at site $x$, i.e., on the states $\phi_{\vec
  n}$ \eqref{eq:23} with $0\leq n_x\leq 2J$, and denote $P_J = \prod_x P_{J,x}$, i.e.,
$P_J$ projects onto the states $\phi_{\vec n}$ with $\vec n\in\mathcal{N}_J$, see
\eqref{eq:25}. By identifying $\phi_{\vec n}$ \eqref{eq:24} with $\phi_{\vec n}$
\eqref{eq:23}, it is clear that
\begin{equation*}
  \Hi = P_J \mathcal{F}
\end{equation*}
where $=$ means unitarily equivalent. Under this equivalence, we find that the spin
operators are given by \cite{hemmen:1984}:
\begin{align}
  \frac 1{\sqrt{2J}} \tilde S^-_x = P_J a^*_x g_J(x)^{\frac 12} 
  && \frac 1{\sqrt{2J}} \tilde S^+_x = g_J(x)^{\frac 12} a_x P_J 
  && J-\tilde S^3_x = P_J a^*_x a_x P_J \label{eq:26}
\end{align}
where
\begin{equation*}
  g_J(x) = g_J(a^*_x a_x)
\end{equation*}
and
\begin{equation*}
  g_J(n)=
  \begin{cases}
    1-\frac 1{2J} n & n\leq 2J\\
    0 & n>2J
  \end{cases}
\end{equation*}

\section{The low energy spectrum}
\label{sec:conv-low-spec}

\subsection{Some estimates for the Hamiltonian} 
\label{sec:time-deriv}

We first need the following identities:
\begin{lemma}\label{lem:angles}
  \begin{align}
    \cos\thet_{x-1} + \cos\thet_{x+1}
    &=\epsilon_x \cos\thet_x \label{eq:11}\\
    \Delta^{-1}(\sin\thet_{x-1} + \sin\thet_{x+1})
    &=\epsilon_x \sin\thet_x \label{eq:12}\\
    \sin\thet_x\sin\thet_{x\pm 1} + \Delta^{-1}\cos\thet_x
    \cos\thet_{x\pm 1}&=\Delta^{-1} \label{eq:13}\\
    \Delta^{-1} \cos\thet_x \sin\thet_{x\pm 1} - \sin\thet_x \cos\thet_{x\pm 1} &=
    \mp\sqrt{1-\Delta^{-2}}\sin\thet_x
    \label{eq:14}\\
    \frac1\Delta \sin\thet_{x\pm 1}\sin\thet_x + \cos\thet_{x\pm 1} \cos\thet_x \mp
    \sqrt{1-\Delta^{-2}} \cos\thet_x &= \epsilon^\pm_x\label{eq:15}
  \end{align}
\end{lemma}
\begin{proof}
  These are straightforward computations using the definitions \eqref{eq:21} and
  \eqref{eq:22} of $\cos\thet_x$ and $\sin\thet_x$, and the addition laws for $\sinh$ and
  $\cosh$.
\end{proof}

With this lemma we can write the Hamiltonian in terms of the $\tilde S$-operators.
\begin{corollary}\label{cor:rotated-ham}
  \begin{align*}
    H^J_{x,x+1} =& J^2 - \frac 1{2\Delta}\bigl( \tilde S^+_x\tilde S^-_{x+1} + \tilde
    S^-_x\tilde S^+_{x+1}\bigr) - \gamma_{x,x+1}
    \tilde S^3_x \tilde S^3_{x+1}\\
    &+J \sqrt{1-\Delta^{-2}}\bigl(
    \cos\thet_x \tilde S^3_x - \cos\thet_{x+1}\tilde S^3_{x+1}\bigr)\\
    &+ \sqrt{1-\Delta^{-2}}\bigl( \sin\thet_x \tilde S^1_x \tilde
    S^3_{x+1} - \sin\thet_{x+1} \tilde S^3_x \tilde S^1_{x+1}\bigr) \\
    &- J \sqrt{1-\Delta^{-2}}\bigl(\sin\thet_x\tilde S^1_x - \sin\thet_{x+1} \tilde
    S^1_{x+1}\bigr)
  \end{align*}
  where
  \begin{equation*}
    \gamma_{x,x+1} = \epsilon_x^+ + \sqrt{1-\Delta^{-2}} \cos\thet_x =
    \epsilon_{x+1}^- - \sqrt{1-\Delta^{-2}} \cos\thet_{x+1}
  \end{equation*}
\end{corollary}
\begin{proof}
  This follows immediately from the relations \eqref{eq:28} -- \eqref{eq:30}, and the
  previous lemma.
\end{proof}

With the Hamiltonian in terms of the $\tilde S$-operators, we can apply the unitary
transformation \eqref{eq:26} to write the spin Hamiltonian as an operator on
$\mathcal{F}$:
\begin{align}\label{eq:39}
  \tfrac1J H_{J,\Lambda} &= P_J\biggl\{ \sum_{x=a}^{b-1}\Bigl[ \epsilon_x^+ g_J(x)
  a^*_x a_x - \Delta^{-1} a^*_{x+1} g_J(x+1)^{\hlf}g_J(x)^{\hlf} a_x
  \Bigr]\nonumber\\ 
  & +\sum_{x=a+1}^{b}\Bigl[ \epsilon_x^- g_J(x) a^*_x a_x - \Delta^{-1} a^*_{x-1}
  g_J(x-1)^{\hlf} g_J(x)^{\hlf}  a_x \Bigl]\nonumber\\
  &+ \frac{\sqrt{1-\Delta^{-2}}}{2J}\Bigl[ \cos\thet_b N_b^2 - \cos\thet_a N_a^2\Bigr] +
  \sum_{x=a}^{b-1} \gamma_{x,x+1}  \frac{(N_x-N_{x+1})^2}{2J}\nonumber\\
  &+\frac{\sqrt{1-\Delta^{-2}}}{2J^{\hlf}} \sum_{x=a}^{b-1}\Bigl[ \sin\thet_{x+1}
  (g_J(x+1)^{\hlf} a_{x+1} + a^*_{x+1}
  g_J(x+1)^{\hlf})   N_{x}\nonumber\\
  &\qquad- \sin\thet_{x} ( g_J(x)^{\frac12 }a_{x} + a^*_{x} g_J(x)^{\hlf}) N_{x+1}
  \Bigr] \biggr\} P_J
\end{align}
and likewise for the $\infty$-volume GNS Hamiltonian:
\begin{align}\label{eq:7}
  \tfrac1J H_J &= P_J\biggl\{ \sum_{x\in\Z}\Bigl[ \epsilon_x g_J(x) a^*_x a_x -
  \Delta^{-1} a^*_x g_J(x)^{\hlf} \bigl( g_J(x-1)^{\scriptscriptstyle{1/2}} a_{x-1} +
  g_J(x+1)^{\hlf} a_{x+1}     \bigr)\Bigr]\nonumber\\
  &+\sum_{x=a}^{b-1} \gamma_{x,x+1} \frac{(N_x-N_{x+1})^2}{2J}\nonumber\\
  &+\frac{\sqrt{1-\Delta^{-2}}}{2J^{\hlf}} \sum_{x\in\Z} \sin\thet_{x} (
  g_J(x)^{\frac12 }a_{x} + a^*_{x} g_J(x)^{\hlf})(N_{x-1}- N_{x+1} )\Bigr] \biggr\} P_J
\end{align}

In the usual language of spin wave theory \cites{dyson:1956,dyson:1956b,hemmen:1984}, the
first term in the Hamiltonian is called the kinematical interaction $\Hkin$, and the
second term the dynamical interaction $\Hdyn$. The last term, which describes transitions
between subspaces with constant number of particles, is not usually present.  We denote it
$\Htran$. Note that we define these three operators with the right scaling already
included, i.e.,
\begin{equation*}
  \tfrac1J H_J = P_J\left\{ \Hkin + \Hdyn +  \Htran \right\} P_J
\end{equation*}

We will only let these operators act on vectors in $\Hi$, hence we may forget about the
$P_J$.  To further simplify some notation, introduce
\begin{itemize}
\item the column vector $A$ of annihilation operators:
  \begin{align*}
    A=\begin{pmatrix}
      \vdots\\ a_x \\ \vdots
    \end{pmatrix}
  \end{align*}
\item the diagonal matrix $G_J$:
  \begin{equation*}
    G_J(x,y) = g_J(x) \delta_{x,y}
  \end{equation*}
\end{itemize}
Then we can write
\begin{align*}
  \Hkin  &= A^* G_J^{\hlf} \tilde h^{(r)} G_J^{\hlf} A - \frac1{2J}
  \sum_x \epsilon_x a^*_x  a_x \\
  \Hr  &= A^* \tilde h^{(r)} A
\end{align*}
where $\hr$ is the one-particle boson Hamiltonian, see \eqref{eq:4}, i.e., the matrix with
entries
\begin{equation*}
  \tilde h^{(r)}(x,y) = \epsilon_x \delta_{x,y} - \Delta^{-1} (\delta_{x-1,y} +
    \delta_{x+1,y}) 
\end{equation*}

In the following we will fix for every $J$ an $n_J\in\N$ with $0<n_J<J$ and make
statements about the subspace $P_{n_J}\Hi$ of $\Hi$. In the end we will formulate results
on the whole of $\Hi$ by adding an external pinning field which will take care of the
states in $(P_{n_J}\Hi)^\perp\cap \Hi$. For simplicity denote $\Hi[n_J] = P_{n_J}\Hi$.

\begin{lemma}\label{lem:bounds-H_kin}
  On $\Hi[n_J]$ we have the lower bound
  \begin{equation*}
    \Hkin \geq (\gapr g_J(2n_J) -\tfrac 1J) \Ntot - \gapr A^*
    G^{\hlf}P_0 G^{\hlf} A
  \end{equation*}
  where $\gapr$ is the spectral gap of $\tilde h^{(r)}$ and $P_0$ is defined in
  \eqref{eq:27}. An upper bound (on the whole $\Hi$) is given by
  \begin{equation*}
    \Hkin\leq \|\tilde h^{(r)}\| \Ntot
  \end{equation*}
\end{lemma}
\begin{proof}
  $P_{n_J}$ projects onto the vectors with at most $2n_J$ particles per site, such that on
  $\Hi[n_J]$:
  \begin{equation*}
    G_J \geq g_J(2n_J)\I
  \end{equation*}
  Obviously $G_J\leq\I$ on $\Hi$. The Lemma follows from the bounds on $\tilde h^{(r)}$:
  \begin{equation*}
    \gapr (\I-P_0) \leq \tilde h^{(r)} \leq \|\tilde h^{(r)}\| \I
  \end{equation*}
  and also $\epsilon_x\leq 2$.
\end{proof}

We are going to compare the spectrum of $\Hkin$ with the spectrum of $\Hr $.
Both operators commute with $\Ntot$ so it is sufficient to compare them on eigenstates of
$\Ntot$. Also for $\Hdyn$ it is sufficient to look at eigenstates of $\Ntot$.

In the proof of the following lemmata we will use the following notation: for $\vec n,
\vec m \in\mathcal{N}$:
\begin{align*}
  T_x^\pm \vec n &= \vec m\;\text{ iff }\;
  \begin{cases}
    m_x = n_x+ 1& \\
    m_{x\pm1} = n_{x\pm1}- 1& \\
    m_y = n_y & \forall y\not= x,x+1
  \end{cases}\\
  A_x^\pm\vec n &= \vec m \;\text{ iff }\; 
  \begin{cases}
    m_x = n_x\pm1 &\\
    m_y = n_y & \forall y\not=x
  \end{cases}
\end{align*}

\begin{lemma}\label{lem:estim-H_kin-H_dyn}
  Let $\psi_N\in \Hi[n_J]$, $\|\psi_N\|=1$, $\Ntot \psi_N= N\psi_N$. Then
  \begin{align*}
    \bigl\|  \bigl( \Hr  - \Hkin\bigr)\psi_N \bigr\| &\leq 2(1+\Delta^{-1})\frac{n_JN}{J}\\
    \bigl\| \Hdyn \psi_N \bigr\| &\leq \frac{4n_J N}{J}
  \end{align*}
\end{lemma}
\begin{proof}
  We can write
  \begin{equation*}
    \psi_N = \sum_{\vec n} c_{\vec n} \;\phi_{\vec n} \text{ with } \sum_{\vec n} |c_{\vec n}|^2=1
  \end{equation*}
  where the sum runs over $\vec n\in\mathcal{N}_J$ for which $\sum_x n_x =N$. On basis
  vectors, we have
  \begin{align*}
    \Hkin \phi_{\vec n} &= \sum_x \epsilon_x n_x g_J(n_x) \phi_{\vec n}\\
    &\quad-\Delta^{-1}\sum_x \bigl[ (n_x+1) g_J(n_x) g_J(n_{x-1}-1) n_{x-1}\bigr]^{\hlf}
    \phi_{T_x^-\vec n}\\
    &\quad-\Delta^{-1}\sum_x \bigl[ (n_x+1) g_J(n_x) g_J(n_{x+1}-1) n_{x+1}\bigr]^{\hlf}
    \phi_{T_x^+\vec n}
  \end{align*}
  and for $\Hr$ the same with the $g_J\equiv 1$. We compare term by term. The first one
  gives:
  \begin{align*}
    \Bigl\| \sum_{\vec n,x} c_{\vec n} \epsilon_x \bigl[n_x -n_x g_J(n_x)\bigr] \phi_{\vec
      n}\Bigr\|^2 &= \frac1{(2J)^2} \sum_{\vec n} |c_{\vec n}|^2 \Bigl| \sum_x \epsilon_x
    n_x^2\Bigr|^2 \leq \frac{4N^2 n_J^2}{J^2}
  \end{align*}
  where we used $\epsilon_x\leq2$, $n_x\leq 2n_J$ and $\sum_x n_x=N$. For the second term
  we use analogously $g_J(n_x)\geq g_J(2n_J)$, and find:
  \begin{align*}
    &\Bigl\|\sum_{\vec n,x} c_{\vec n} \Bigl( \bigl[ (n_x+1)n_{x-1}\bigr]^{\hlf} - \bigl[
    (n_x+1) g_J(n_x) g_J(n_{x-1}-1) n_{x-1}\bigr]^{\hlf}\Bigr) \phi_{T_x^-\vec n}\Bigr\|\\
    &=\Bigl\|\sum_{\vec m,x} c_{T_{x-1}^+\vec m} \Bigl( \bigl[ m_x
    (m_{x-1}+1)\bigr]^{\hlf} - \bigl[ m_x g_J(m_x-1) g_J(m_{x-1})
    (m_{x-1}+1)\bigr]^{\hlf}\Bigr) \phi_{\vec m}\Bigr\|\\
    &\leq\sum_{\vec m}\Bigl( \sum_x |c_{T_{x-1}^+\vec m}| \bigl| \bigl[ m_x
    (m_{x-1}+1)\bigr]^{\hlf} - \bigl[ m_x g_J(m_x-1) g_J(m_{x-1})
    (m_{x-1}+1)\bigr]^{\hlf}|\Bigr)^2\\
    &\leq \sum_{\vec m}\Bigl( \sum_x |c_{T_{x-1}^+\vec m}| \bigl[ m_x
    (m_{x-1}+1)\bigr]^{\hlf} (1-g_J(2n_J)) \Bigr)^2\\
    &\leq \frac{n_J^2}{J^2}\sum_{\vec m} \Bigl( \sum_x |c_{T_{x-1}^+\vec m}|^2 (m_{x-1}+1)
    \Bigr) \Bigl( \sum_x m_x\Bigr) = \frac{N^2 n_J^2}{J^2}
  \end{align*}
  and the same for the third term. Summing everything together we find
  \begin{equation*}
    \bigl\| (\Hr - \Hkin)\psi_N\bigr\| \leq 2(1+\Delta^{-1})\frac{N n_J}{J}
  \end{equation*}
  For $\Hdyn$ we use the same reasoning,  $\gamma_{x,x+1}\leq 1$, and 
  \begin{equation*}
    (n_x-n_{x+1})^2 \leq 2 n_x^2 + 2 n_{x+1}^2 \leq 4n_J( n_x + n_{x+1})
  \end{equation*}
  to find
  \begin{equation*}
    \bigl\| \Hdyn \psi_N\bigr\| \leq \frac{4n_J N}{J}
  \end{equation*}
\end{proof}

For $\Htran$ we have the following estimate.
\begin{lemma}\label{lem:estim-H_tran}
  Let $\psi\in \Hi[n_J]$, $\|\psi\|=1$. Then
  \begin{align*}
    \bigl\| \Htran \psi\bigr\| &\leq 2\sqrt{1-\Delta^{-2}} |v_0|_1
    \Bigl(\frac{(2n_J+1) (4n_J)^2}{J}\Bigr)^{\hlf}
  \end{align*}
  where $|v_0|_1$ is the $l^1$-norm of the zero-mode.
\end{lemma}
\begin{proof}
  Let
  \begin{equation*}
    \psi=\sum_{\vec n} c_{\vec n}\phi_{\vec n}
  \end{equation*}
  Then
  \begin{align*}
    &\Bigl\|\sum_x \sin\thet_x g_J(x)^{\hlf} a_x (N_{x-1}-N_{x+1}) \psi\Bigr\|^2\\
    &\quad =\Bigl\| \sum_{\vec n,x} c_{\vec n} \sin\thet_x g_J(n_x-1)^{\hlf}
    n_x^{\hlf}    (n_{x-1} - n_{x+1}) \phi_{A_x^-\vec n}\Bigr\|^2\\
    &\quad =\Bigl\| \sum_{\vec n,x} c_{A_x^+\vec n} \sin\thet_x g_J(n_x)^{\hlf}
    (n_x+1)^{\hlf}  (n_{x-1} - n_{x+1}) \phi_{\vec n}\Bigr\|^2\\
    &\quad= \sum_{\vec n} \Bigl| \sum_x c_{A_x^+\vec n} \sin\thet_x g_J(n_x)^{\hlf}
    (n_x+1)^{\hlf} (n_{x-1} - n_{x+1}) \Bigr|^2\\
    &\quad\leq \sum_{\vec n} \bigl(\sum_x \sin\thet_x |c_{A_x^+\vec n}|^2 \bigr) \bigl(
    \sum_x \sin\thet_x (n_x+1) |n_{x-1} - n_{x+1}|^2\bigr)\\
    &\quad\leq |v_0|_1 (2n_J+1) (4n_J)^2 \sum_{\vec n} \sum_x \sin\thet_x |c_{A_x^+\vec
      n}|^2\\
    &\quad= |v_0|^2_1 (2n_J+1) (4n_J)^2
  \end{align*}
  and likewise for the second term.
\end{proof}

\subsection{Strong convergence and spectral concentration}
\label{sec:strong-conv}

Recall the following definitions.
For $A$ a self-adjoint operator and $a,b\in\R$, denote by $P_{(a,b)}(A)$ the spectral
projection of $A$ onto $(a,b)$. For $A$ acting on Fock space, denote
\begin{itemize}
\item $\sigma(A)$ the spectrum of $A$ in $\mathcal{F}$;
\item $\sigma_J(A)$ the spectrum of $P_JAP_J$ in $\Hi$;
\item $\sigma_{n_J}(A)$ the spectrum of $P_{n_J}AP_{n_J}$ in $\Hi[n_J]$.
\end{itemize}
Also denote $\slim$ the strong, or strong resolvent, operator limit for bounded,
resp. unbounded operators acting on $\mathcal{F}$.

In the GNS space $\Hi$ it is convenient to define $\Stot[]$ in the renormalized sense: 
\begin{equation*}
  \Stot[] = \sum_{x\in\Z}\bigl[ S^3_x - \sgn(x-\tfrac12)\bigr]
\end{equation*}
Also denote
\begin{equation*}
  \mu=\sum_{x\in\Z} \bigl[\cos\thet_x - \sgn(x-\tfrac12)\bigr]
\end{equation*}
i.e., $\mu$ is the $3$-magnetization of a classical ground state
$\{(\thet_x,\phi)\}_{x\in\Z}$ (see section \ref{sec:xxz-class-lim}).

\begin{proposition}\label{pro:strong-convergence}
  We have
  \begin{align*}
    \slim_{J\to\infty} \tfrac1J H_J &= \Hr\\
    \slim_{J\to\infty} \tfrac1J \Stot[] &= \mu \I
  \end{align*}
\end{proposition}
\begin{proof}
  Introduce the set $\mathcal{D}$, the finite linear space of vectors $\phi_{\vec n}$ with
  $\vec n\in\mathcal{N}$. $\mathcal{D}$ is a common core for $\tfrac 1J H_J$, for all $J$,
  and $\Hr $.
  
  Take $\psi\in\mathcal{D}$ arbitrary (but normalized for simplicity) and denote
  \begin{align*}
    \psi &= \sum_{\vec n} c_{\vec n} \phi_{\vec n}\\
    N_\psi &= \sup_{\vec n\colon c_{\vec n}\not= 0} \sum_x n_x
  \end{align*}
  Note that by assumption, $\psi$ is a finite sum of $\phi_{\vec n}$ with $\sum_x
  n_x<\infty$, and hence also $N_\psi<\infty$.
  
  Now take $J$ large enough such that $\psi\in\Hi$. From the proof of Lemma
  \ref{lem:estim-H_kin-H_dyn} and \ref{lem:estim-H_tran} it is clear that we can use
  $2n_J\leq N_\psi$, as soon as $2J>N$, hence
  \begin{equation*}
    \bigl\| (\tfrac 1J H_J -\Hr )  \psi\bigr\|\leq (3+\Delta^{-1})\frac{N_\psi^2}J + 2
    \sqrt{1-\Delta^{-2}} |v_0|_1 \Bigl( \frac{(N_\psi+1)(2N_\psi)^2}J\Bigr)^{\hlf}
  \end{equation*}
  The first result follows from \cite[Theorem VIII.25 (a)]{reed:1972}.

  To prove the second statement, write
  \begin{align*}
    S^3_x &= \cos\thet_x \tilde S^3_x - \sin\thet_x \tilde S^1_x\\
    &= P_J\Bigl\{ \cos\thet_x (J-N_x) - \sin\thet_x \bigl( g_J(x)^{\frac 12}
     a_x + a^*_x
    g_J(x)^{\frac 12}\bigr) \Bigr\} P_J
  \end{align*}
  and hence
  \begin{align*}
    \Stot[]&=\sum_x S^3_x - \sgn(x-\tfrac12)\\
    &= P_J\Bigl\{ \mu J - \sum_x \cos\thet_x N_x - \sum_x \sin\thet_x 
    \bigl( g_J(x)^{\frac 12} a_x +
    a^*_x g_J(x)^{\frac 12}\bigr) \Bigr\} P_J
  \end{align*}
  Clearly $\sum_x\cos\thet_x N_x\leq \Ntot$, and for $\psi\in\mathcal{D}$ as before
  \begin{align*}
    \Bigl\|\sum_x \sin\thet_x a^*_x g_J(x)^{\frac 12} 
    \bigl( \sum_{\vec n} c_{\vec n} \phi_{\vec
      n}\bigr) \Bigr\|^2 &= \Bigl\|\sum_{\vec n,x} \sin\thet_x c_{\vec n}
    (n_x+1)^{\hlf}
    g_J(n_x)^{\frac 12} \phi_{A_x \vec n}\Bigr\|^2\\
    &\leq (N_\psi +1) |v_0|_1^2
   \end{align*}
   Hence
   \begin{align*}
     \Bigl\| \bigl(\tfrac 1J \Stot[] - \mu\bigr) \psi\Bigr\| &\leq \frac 1J\bigl( N_\psi +
     2 (N_\psi +1)^{\hlf} |v_0|_1 \bigr)
   \end{align*}
   and $\tfrac 1J \Stot[]\to \mu\Eins$ strongly on $\mathcal{F}$.
\end{proof}

\begin{corollary}\label{cor:strong-convergence}\hspace*{0pt}
  \begin{enumerate}
  \item If $\lambda\in\sigma(\Hr )$, there exists $\lambda_J\in\sigma_J(\tfrac1JH_J)$ such
    that
    \begin{equation*}
      \lim_{J\to\infty} \lambda_J = \lambda
    \end{equation*}
  \item If $a,b\in\R$, and $a,b\notin\sigma_{pp}(\Hr )$, then 
    \begin{equation*}
      \slim_{J\to\infty}  P_{(a,b)}(\tfrac1J H_J) =  P_{(a,b)}(\Hr)
    \end{equation*}
  \item If $a,b\in\R$, and $\mu\in (a,b)$, then
    \begin{equation*}
      \slim_{J\to\infty} P_{(a,b)}(\tfrac1J\Stot[]) = \I
    \end{equation*}
  \end{enumerate}
\end{corollary}
\begin{proof}
  The previous proposition and \cite[Theorem VIII.24]{reed:1972}.
\end{proof}

In addition we can prove spectral concentration of $\tfrac1J H_J$ around discrete
eigenvalues of $\Hr $.
\begin{proposition}\label{pro:spectr-conc}
  For every isolated eigenvalue $E$ of $\Hr $, there exists an interval
  \begin{align*}
    I_J=(E-\epsilon_J, E+\epsilon_J)\quad\text{ with }\quad
    \lim_{J\to\infty} \epsilon_J\frac{J}{\ln J} = 0
  \end{align*}
  such that for any interval $I$ around $E$ s.t. $I\cap \sigma(\Hr) = \{E\}$:
  \begin{align*}
    \slim_{J\to\infty} P_{I\setminus I_J}\bigl(\tfrac 1J H_J\bigr) &= 0\\
    \slim_{J\to\infty} P_{I_J} \bigl(\tfrac 1J H_J\bigr)
    &= P_{\{E\}} \bigl(\Hr  \bigr)
  \end{align*}
\end{proposition}
\begin{proof}
  Let $\psi_E$ be the simultaneous eigenvector of $\Hr $ with eigenvalue $E$,
  and of $\Ntot$ with eigenvalue $N_E$. It follows that 
  \begin{equation*}
    N_E \leq \frac E{\gapr}
  \end{equation*}
  We have from Lemma \ref{lem:estim-H_kin-H_dyn}
  \begin{equation*}
    \bigl\|( \tfrac 1J H_J - E) \psi_E\bigr\| \leq \frac{c(\Delta,r,E)}J
  \end{equation*}
  or, 
  \begin{equation*}
    \lim_{J\to\infty} \frac J{\ln J}\bigl\|( \tfrac 1J H_J - E) \psi_E\bigr\|=0
  \end{equation*}
  Hence it follows that $E$ is a first order pseudo-eigenvalue with first-order pseudo
  eigenvector $\psi_E$, and the result follows from \cite[Theorem XII.22]{reed:1972}.
\end{proof}

\begin{remark}
  Applying the same reasoning to $\Stot[]$, we find that the interval $(a,b)$ in item
  (iii) of Corollary \ref{cor:strong-convergence} can be chosen as
  \begin{equation*}
    (a,b) =  (\mu-\epsilon_J, \mu+\epsilon_J)
  \end{equation*}
  with again $\lim_J \epsilon_J J(\ln J)^{-1}=0$.
\end{remark}

\subsection{Convergence of the spectrum with a pinning field}
\label{sec:with-pinning}

To prove full convergence of the spectrum, we have to add the external field
\eqref{eq:17} to $H_J$, or, to have a positive operator, add:
\begin{equation}\label{eq:32}
  F_J=\sup_x \bigl( |h_{J,x}| J - \vec h_{J,x}\cdot \vec S_x \bigr)
  = h_J \sup_x (J-\tilde S^3_x) = h_J \sup_x (N_x)
\end{equation}
with $h_J>0$, see also the end of section \ref{sec:gc-states}, and $N_x=a^*_x a_x$. Let us
assume we add this field to $\tfrac 1J H^J$, so $h_J$ already contains the factor
$J^{-1}$.

Take $0<n_J<J$ as before, we get that on $\Hi[n_J]^\perp\cap\Hi$
\begin{equation*}
  \tfrac 1J H_J + h_J \sup_x N_x \geq h_J \sup_x N_x \geq h_J n_J \I
\end{equation*}
Clearly, by choosing $h_J$ such that
\begin{equation*}
  \lim_{J\to\infty} h_J n_J =\infty
\end{equation*}
statements about the spectrum on $\Hi$ reduce to statements about the spectrum on
$\Hi[n_J]$. Or, if one chooses to make statements about the spectrum below a certain value
$E$, it is sufficient to choose $h_J$ such that $\lim_J h_J n_J > E+\epsilon$.

Convergence of the spectrum of $\Hkin + \Hdyn$ can be proved under the weakest assumptions
on $n_J$. First we prove convergence of the spectrum of $\Hkin$.

\begin{proposition}
  Let $n_J= [ J(\ln J)^{-1} ]$, where $[\cdot]$ denotes the integer part.  If
  $\lambda\notin \sigma(\Hr)$, then $\lambda\notin \sigma_{n_J}(\Hkin)$ for $J$ large
  enough.
\end{proposition}
\begin{proof}
  Assume that $\lambda\in \sigma_{n_J}(\Hkin)$ for all $J$ larger than some
  $J_0$. Take $\delta>0$ arbitrary, and $\psi_J\in\mathcal{H}_{n_J}$ a
  normalized approximate eigenvector:
  \begin{equation*}
    \bigl\| (\Hkin - \lambda)\psi_J\bigr\|<\delta
  \end{equation*}
  More precisely, take $\psi_J$ such that
  \begin{equation*}
    P_{(\lambda-\delta,\lambda+\delta)}(\Hkin)\psi_J = \psi_J
  \end{equation*}
  Since $[\Hkin,\Ntot]=0$ we can take $\psi_J$ an eigenstate of $\Ntot$ with eigenvalue
  $N_J$. Since $\psi_J$ is orthogonal to the ground state space, it follows from Lemma
  \ref{lem:bounds-H_kin} that
  \begin{equation*}
    N_J \leq \frac{\lambda+\delta}{\gapr g_J(2n_J) - J^{-1}}
  \end{equation*}
  By our choice of $n_J$, we have $\lim_J n_J J^{-1}=0$, or $\lim_J g_J(2n_J)=1$, and for
  $J$ large enough,
  \begin{equation*}
    N_J \leq \frac{2(\lambda+\delta)}{\gapr}
  \end{equation*}
  Putting this into the bounds of Lemma \ref{lem:estim-H_kin-H_dyn}, we find
  \begin{align*}
    \bigl\|(\Hr -\lambda)\psi_J\bigr\| &\leq \bigl\|(\tilde
    H^{(r)}-\Hkin)\psi_J\bigr\| + \bigl\| (\Hkin -  \lambda)\psi_J\bigr\|\\
    &\leq \frac{4(1+\Delta^{-1})(\lambda+\delta)}{\gapr}\frac{n_J}J +\delta
  \end{align*}
  and it follows that $\psi_J$ is an approximate eigenvector for $\Hr $ as well and
  $\lambda\in\sigma(\Hr )$.
\end{proof}

Now we add $\Hdyn$:
\begin{proposition}
  Let again $n_J= [ J(\ln J)^{-1} ]$. If $\lambda\notin \sigma(\Hr )$, then $\lambda\notin
  \sigma_{n_J}(\Hkin+\Hdyn)$ for $J$ large enough.
\end{proposition}
\begin{proof}
  Since $\Hdyn\geq 0$ it follows that an approximate eigenvector $\psi_J$ for
  $\Hkin+\Hdyn$ must satisfy
  \begin{equation*}
    \bigl\|\Hkin \psi_J\bigr\| \leq \lambda +\delta
  \end{equation*}
  with the same notation as in the previous proposition. Hence we get the same estimate on
  $N_J$ as before, but this implies by Lemma \ref{lem:estim-H_kin-H_dyn} that $\lim_J
  \|\Hdyn \psi_J\|=0$ and $\psi_J$ is an approximate eigenvector for $\Hr $ as
  well.
\end{proof}

Alternatively, these propositions prove that the spectrum of $\Hkin +\Hdyn + h_J \sup_x
N_x$ converges to the spectrum of $\Hr$, provided
\begin{align*}
  \lim_J h_J=0 && \lim_J \frac{h_JJ}{\ln J} = \infty
\end{align*}

To add $\Htran$ we clearly have to relax our condition on $n_J$.
\begin{proposition}
  Let $n_J = [ (J (\ln J)^{-1})^{\thrd}]$.  If $\lambda\notin \sigma(\Hr )$, then
  $\lambda\notin \sigma_{n_J}(\tfrac 1J H_J)$ for $J$ large enough.
\end{proposition}
\begin{proof}
  By Lemma \ref{lem:estim-H_tran} we have for all $\psi\in\mathcal{H}_{n_J}$
  \begin{equation*}
    \| \Htran \psi\| \leq 2\sqrt{1-\Delta^{-2}} |v_0|_1
    \Bigl(\frac{(2n_J+1) (4n_J)^2}{J}\Bigr)^{\hlf} \|\psi\|
  \end{equation*}
  and by assumption the r.h.s. goes to $0$ as $J\to\infty$.
\end{proof}

Hence we get convergence of the spectrum of $\tfrac 1J H_J + h_J \sup_x N_x$ if
\begin{align*}
  \lim_{J\to\infty} h_J =0 && \lim_{J\to\infty} h_J\Bigl(\frac{J}{\ln J}\Bigr)^{\thrd} =\infty
\end{align*}

The results obtained in this section include the statements made in 
Section \ref{sec:statements}.

%




\end{document}